\documentclass[acmsmall]{acmart}
\usepackage{multirow}

\usepackage{amsthm}
\usepackage{graphicx}
\usepackage{graphics}
\usepackage{subfigure}
\usepackage{amsmath}
\usepackage{float}
\usepackage{algorithm}     
\usepackage{algorithmic}

\AtBeginDocument{%
  \providecommand\BibTeX{{%
    \normalfont B\kern-0.5em{\scshape i\kern-0.25em b}\kern-0.8em\TeX}}}

\acmJournal{JACM}
\acmVolume{37}
\acmNumber{4}
\acmArticle{111}
\acmMonth{8}

\begin{document}

\title{Disentangled Item Representation for Recommender Systems}

\author{Zeyu Cui}
\email{zeyu.cui@nlpr.ia.ac.cn}
\orcid{0000-0003-0017-5292}
\affiliation{%
  \institution{Institution of Automation Chinese Academy of Sciences (CASIA)}
  \institution{University of Chinese Academy of Sciences (UCAS)}
  \institution{National Laboratory of Pattern Recognition (NLPR)}
  \institution{Center for Research on Intelligent Perception and Computing(CRIPAC)}
  \streetaddress{No.95 ZhongGuanCun East Road}
  \city{Beijing}
  \postcode{100080}
}

\author{Feng Yu}
\email{yf271406@alibaba-inc.com}
\affiliation{%
  \institution{Alibaba Group}
  \institution{Institution of Automation Chinese Academy of Sciences (CASIA)}
  \institution{University of Chinese Academy of Sciences (UCAS)}
  \institution{National Laboratory of Pattern Recognition (NLPR)}
  \institution{Center for Research on Intelligent Perception and Computing(CRIPAC)}
  \streetaddress{No.95 ZhongGuanCun East Road}
  \city{Beijing}
  \postcode{100080}
}

\author{Shu Wu}
\authornote{Corresponding author.}
\email{shu.wu@nlpr.ia.ac.cn}
\author{Qiang Liu}
\email{qiang.liu@nlpr.ia.ac.cn}
\author{Liang Wang}
\email{wangliang@nlpr.ia.ac.cn}
\affiliation{%
  \institution{Institution of Automation Chinese Academy of Sciences (CASIA)}
  \institution{University of Chinese Academy of Sciences (UCAS)}
  \institution{National Laboratory of Pattern Recognition (NLPR)}
  \institution{Center for Research on Intelligent Perception and Computing(CRIPAC)}
  \streetaddress{No.95 ZhongGuanCun East Road}
  \city{Beijing}
  \postcode{100080}
}

%
%
%
%

\renewcommand{\shortauthors}{Cui, et al.}

\begin{abstract}
  Item representations in recommendation systems are expected to reveal the properties of items.
Collaborative recommender methods usually represent an item as one single latent vector.
Nowadays the e-commercial platforms provide various kinds of attribute information for items (e.g., category, price and style of clothing). Utilizing these attribute information for better item representations is popular in recent years.
Some studies use the given attribute information as side information, which is concatenated with the item latent vector to augment representations. 
However, the mixed item representations fail to fully exploit the rich attribute information or provide explanation in recommender systems.
To this end, we propose a fine-grained Disentangled Item Representation (DIR) for recommender systems in this paper, where the items are represented as several separated attribute vectors instead of a single latent vector.
In this way, the items are represented at the attribute level, which can provide fine-grained information of items in recommendation. 
We introduce a learning strategy, LearnDIR, which can allocate the corresponding attribute vectors to  items.
We show how DIR can be applied to two typical models, Matrix Factorization (MF) and Recurrent Neural Network (RNN). Experimental results on two real-world datasets show that the models developed under the framework of DIR are effective and efficient. 
Even using fewer parameters, the proposed model can outperform the state-of-the-art methods, especially in the cold-start situation. 
In addition, we make visualizations to show that our proposition can provide explanation for users in real-world applications.
\end{abstract}



\setcopyright{acmlicensed}
\acmJournal{TIST}
\acmYear{2020} \acmVolume{1} \acmNumber{1} \acmArticle{1} \acmMonth{1} \acmPrice{15.00}\acmDOI{10.1145/3445811}

\maketitle

\section{Introduction}
\noindent The item representations are significant in recommendation systems, which are expected to reveal the item properties.
Early collaborative recommender methods represent an item as a single latent vector only using collaborative information \cite{rendle2009bpr,koren2009matrix,zhang2019deep}.
Today's e-commercial platforms provide various kinds of attribute information for items (e.g., category, price and style of clothing) helping customers select their favorite items. 
These attribute information of items is significant but have not been fully exploited in recommender systems. 
Utilizing these information for better recommendation is preferred in recent years. 
For example, based on traditional collaborative methods, researchers try to add more attribute information to augment the representations \cite{lian2014geomf,he2016vbpr},  
by simply concatenating the attribute information with the latent item vectors and learning item representations by user-item interactions.
Recently, some methods on clothing recommendation begin to analyze items from two attributes, i.e., category and style \cite{he2016sherlock,liu2017deepstyle,yu2018aesthetic}. 
Their success inspires us that 
an item could be considered as a combination of different attributes. 
Compared with item-level representations,  attribute-level representations provide a more informative description for items.

\begin{figure}[t]
\centering
\includegraphics[width=0.8\linewidth]{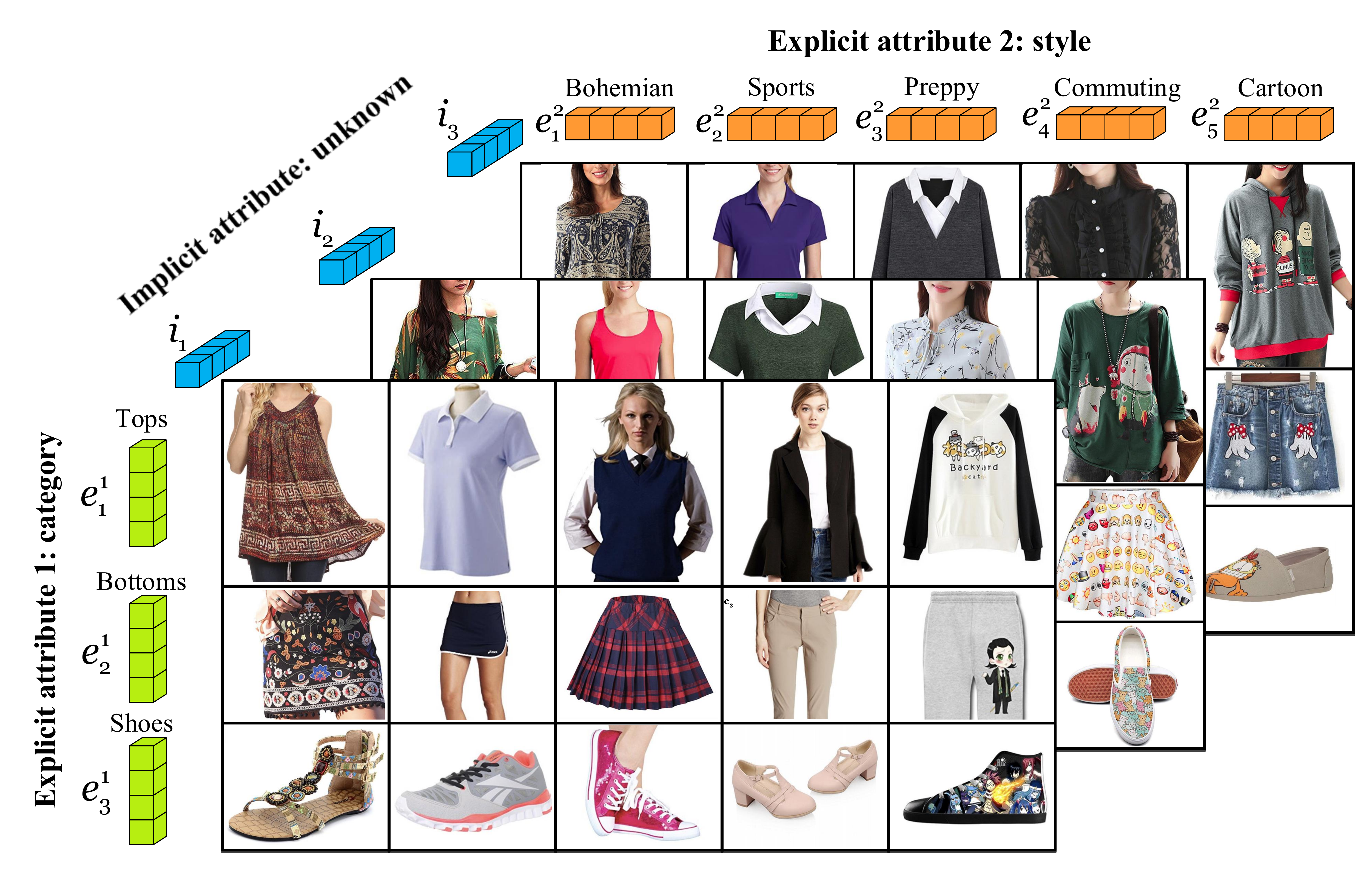}
\caption{
The framework of DIR with 2 given explicit attributes: category and style. 
Each item can be represented as a combination of these three kinds of attributes while each attribute is represented as an attribute vector. The number of implicit attribute vectors is pre-defined.
}
\label{fig:dot}
\end{figure}

In this paper, we formalize an novel item representation framework at the attribute level, Disentangled Item Representation (DIR), as shown in Figure \ref{fig:dot}. Each item is represented as a combination of attribute vectors instead of a single latent vector. 
Aside from the given attributes of items (generally called \emph{explicit attributes}), we introduce an \emph{implicit attribute} to distinguish items from those having the same given attributes.
To better illustrate our framework, we take items of two explicit attributes for example.
Assume the clothes have two kinds of
given attributes, category and style. 
As shown in Figure \ref{fig:dot}, each item can be allocated into a cell of the 3-axes tensor, where each axis corresponds to one kind of attribute (2 explicit attributes and 1 implicit attribute).
Accordingly, we have two main tasks: 
1) assign each item with its corresponding implicit attribute (the explicit attribute are given), i.e., allocate each item into a cell of the tensor;
2) learn all the attribute vectors. 
We also introduce a learning algorithm, LearnDIR,
in which we first randomly initialize all the attribute vectors and allocate them to items, and then we train the vectors and reallocate them to items alternately during training process.

DIR has several advantages compared with traditional methods.
On one hand, DIR reduces the size of parameters.
For a table with $m$ rows and $n$ columns, using $(m+n)$ latent attribute vectors, DIR can represent $(mn)$ items at most.
In pervious traditional models,
the number of latent vectors equals the number of items (i.e., $mn$ in the Figure \ref{fig:dot}) at least.
On the other hand, DIR alleviates the cold-start problem. Cold-start items seldom appear on history records, indicating sparse collaborative information at the item level.
However, DIR models items by shared attributes, where collaborative information is dense.
DIR could be used in most collaborative recommender models. In this paper, we apply DIR on two typical models (i.e., MF and RNN). The proposed models are accordingly called DIR-MF and DIR-RNN.
Experimental results on two real-world datasets show that the models developed under the framework of DIR are effective and efficient. Even using fewer parameters, the proposed model can outperform the state-of-the-art methods. In addition, we make visualizations to show that our proposition can provide the explainability for users in real-world applications.

Overall,  main contributions of our work are summarized as follows:

\begin{itemize}
\item
We propose a novel disentangled item representation for recommender systems, which can fully reveal the properties of items and provide explanation for making recommendation. 
\item 
DIR can reduce the size of parameters. It is a good solution to relieve the cold-start problem by taking advantage of the attribute sharing mechanism.
\item
We show how to implement DIR on two typical models, MF and RNN and demonstrate that DIR can represent items effectively and efficiently. 
Extensive experiments on real-world datasets show that our proposed DIR-RNN model outperforms the state-of-the-art methods. 	
\end{itemize}

\section{Related Work}

\noindent In this section, we briefly review some related works on representation learning and collaborative recommender methods for implicit feedback.

\subsection{Disentangled Representation Learning}
 \noindent The representation learning algorithms are mainly discussed in natural language processing.   The word embedding techniques achieve a great success, such as Word2Vec \cite{mikolov2013efficient,mikolov2013distributed} and Glove \cite{pennington2014glove}. 
Recently, LightRNN \cite{li2016lightrnn} models the items' commonalities based on the 2-component shared embedding, which largely reduces the parameter size. 
This work disentangles the word representations into two components at the first time and achieves considerable performance. 
Many researchers begin to realize the significance of representation disentangling. 
Disentangled representation is commonly the idea of analyzing different attributes of image or text. 
Michael et al \cite{mathieu2016disentangling} and Xi Chen et al \cite{chen2016infogan} try to split the style information from image by representation disentangling.
Vineet et al \cite{john2018disentangled} use the auto-encoder model to disentangle the sentiment information of text. 
Not only in the area about image or text, disentangled representations can capture the descriptive entity \cite{higgins2018towards} of the real world. 
In e-commercial applications, items representation could be disentangled as attributes, which may bring the bonus for recommendation.

\subsection{Item Representation Learning}
 \noindent In recommendation applications, previous researchers enrich the item representation by using more attribute information. 
The work of \cite{lian2014geomf} simply augments the item representation
by concatenating location information with latent factors. 
Sun et al \cite{sunmrlr} uses the information of the category tree to 
give item hierarchical representation.
These works try to augment the item representation by using attribute information to describe items more elaborately. 
Some works also begin to analyze the items' attributes in detail. For example, in clothing recommendation, researchers describe items from the view of category and style \cite{mcauley2015image}. 
He et al \cite{he2016sherlock} and Liu et al \cite{liu2017deepstyle} hold the view that both category and style reflect properties of items together. 
Sherlock \cite{he2016sherlock} uses well-designed linear projections to get style representations from visual information, 
where different categories of items have different projections.
DeepStyle \cite{liu2017deepstyle} declares that style representation could be obtained by splitting category information from visual representation. 
Following the ideas above, some recent works map items into different attribute spaces to get multiple views of items.
 For example, Yu et al \cite{yu2018aesthetic} extract both aesthetic features and category features from item images to represent items in aesthetic space and category space. Sun et al \cite{sun2017exploiting,sunmrlr} try to model the category tree to get different item representations in different  hierarchical spaces. 
 Recently, some researches on fashion recommendation begin to adopt more attribute tags \cite{bao2019collaborative,yang2019interpretable,hou2019explainable} and content information \cite{hu2019transfer,yan2019differentiated,shin2019deep} to enrich items. 
 Different from previous works, our proposed model directly represents items by their attributes. Each item has a disentangled representation at the attribute level.

 \subsection{Collaborative Methods for Implicit Feedback}
 \noindent Implicit feedbacks come from the users' behavior on the internet, which could be tracked automatically, such as clicks, purchases, and so on. Compared with explicit feedbacks like ratings, 
implicit feedbacks are easier to be collected, but more difficult to be modeled because of containing only positive observations.
Collaborative recommender models are mainly used for implicit feedback.
Matrix factorization (MF) \cite{koren2009matrix} is a classical collaborative method. Especially, One-Class MF \cite{pan2008one} treats the non-observed interactions as negative samples. After that, Bayesian Personalized Ranking (BPR) \cite{rendle2009bpr} is introduced, using pair-wise ranking strategies to balance the positive and negative samples. By using the BPR framework, BPR-MF achieves the state-of-the-art performance in many scenarios.
 
Since then, some works try to apply BPR to sequential recommendation. For example, Factorizing Personalized Markov Chains (FPMC) \cite{rendle2010factorizing} uses Markov Chains to model sequential behaviors of users and employs BPR to learn the factorization parameters. Recently, some researchers begin to use RNN to model the sequential behaviors \cite{yu2016dynamic,liu2016predicting}. 
Long Short-Term Memory (LSTM) \cite{hochreiter1997long,gers1999learning,sundermeyer2012lstm} is a popular RNN architecture, which can alleviate the long-term dependence problem of RNN.
Various RNN based models tend to use LSTM for recommendation.


\section{Proposed Method}
 \noindent In this section, 
 we first give the basic notation used in this paper and then propose DIR which could be used for collaborative recommendation models. After that, we introduce LearnDIR for DIR learning. Finally, we show how DIR can be combined with two typical models, MF and RNN.

\subsection{Notation}
 \noindent We use a tuple $(u, q)$ to describe the user behavior, which means that a user $u$ buys an item $q$. Let $\mathcal{U}$ be the set of all users and $\mathcal{Q}$ be the set of all items. 
The purchase set could be formalized as $\mathcal{S} = \{ (u_{1},q_{1}),(u_{2},q_{2})... \}$. 
The task of the recommender algorithms is to provide each user $u$ with a personalized score on each item $q$, denoted as $x_{u, q}$. The larger value of $x_{u, q}$, the higher probability that $u$ buy $q$.
For convenience, we also define:
\[
\mathcal{Q}_{u} = \{ q \in \mathcal{Q} | (u, q) \in \mathcal{S}\},~
\]
\[
\mathcal{U}_{q} = \{ u \in \mathcal{U} | (u, q) \in \mathcal{S}\},~
\]
where $\mathcal{Q}_{u}$ is the set of  items which are purchased by the user $u$, and $\mathcal{U}_{q}$ is the set of users who have purchased the item $q$.

\subsection{Disentangled Item Representation} \label{section_DIR}
 \noindent We formulate that an item $q$ is represented by the combination of several explicit attributes (given) and an implicit attribute (unknown). 
There are $N$ kinds of explicit attributes noted as,
\[
\mathcal{E}^{k} = \{ e^{k}_{1}, e^{k}_{2}, ...  \}, k = \{ 1, 2, ..., N\}.
\]
Each explicit attribute $e$ represents a known attribute.
The implicit attribute is noted as,
\[
\mathcal{I} = \{ i_{1}, i_{2}, ...  \}
\]
implicit attributes is defined to describe the latent attribute of items, which is not described by explicit attributes. By adding implicit attributes, items with the same explicit attributes could be distinguished from each other.
We pre-define the number of implicit attributes for a specific dataset to guarantee each item belongs a unique combination (detail discussion in Section \ref{sec:implicit_num}).
We have two tasks: 1) allocate attributes to items and 2) learn all the attribute vectors $e$ and $i$. 
Mathematically, we use allocating functions $l(q, e)$ (or $l(q, i)$) to formulate whether $q$ belongs to an explicit attribute $e$ (or an implicit attribute $i$).  
$l(q, e)=1$ means that the item $q$ belongs to the attribute $e$, otherwise $l(q, c) =0$, and $l(q, i)$ has the same analogous definition. 
Task 1 could be considered as allocating each item $q$ into a cell of a tensor with $N+1$ axes (i.e., $N$ explicit attributes and $1$ implicit attribute).
In the following, Equation (\ref{constraint1}) makes sure that each cell of the tensor could have at most one item. 
Equation (\ref{constraint2}) guarantees that each item could only be allocated to one cell of the tensor. 
Satisfying both Equations of constraints, each item could be uniquely represented.
\begin{equation} \label{constraint1}
	 \sum_{q \in \mathcal{Q}} l(q, e^{1})l(q, e^{2})...l(q, e^{N})l(q, i) \leq 1, \forall e^{k} \in \mathcal{E^{k}} , \forall i \in \mathcal{I},~
\end{equation}
\begin{equation} \label{constraint2}
	  \sum_{e^{1} \in \mathcal{E}^{1}} 
	   \sum_{e^{2} \in \mathcal{E}^{2}}
	   ...
	   \sum_{e^{N} \in \mathcal{E}^{N}}
	   \sum_{i \in \mathcal{I}}
	  l(q, e^{1})l(q, e^{2})...l(q, e^{N})l(q, i) = 1, \forall q \in \mathcal{Q}.~
\end{equation}	

We denote the preference of user $u$ for explicit attribute $e$ and implicit attribute $i$ as $x_{u, e}$ and $x_{u, i}$.
Given a user behavior tuple ($u$, $q$), instead of predicting $x_{u, q}$ directly, we predict his preference for attributes respectively.
$x_{u, q}$ can be formulated as:
\begin{equation} \label{eqh_ratse}
x_{u, q} = x_{u,e^{1}}x_{u,e^{2}} ... x_{u,e^{N}} x_{u,i}.~
\end{equation}
where item $q$ is represented by $(e^{1}, e^{2}, ..., e^{N}, i)$. Commonly, preference indicts the favor degrees of a user for items. In this paper, we extend the meaning of preference, which indicts the favor degrees of a user for attributes. In that case,  the favor degrees of a user $u$ to an item $q$ (i.e., $x_{u,q}$) could be considered as the aggregation of the preference of $u$ for all the attributes that $i$ possesses. Here, we use multiply as aggregation for simplification.

\subsection{LearnDIR} \label{learnDIR}
 \noindent LearnDIR aims to minimize the negative log loss, 
which is formulated as,
\begin{equation} \label{eqh_obj}
\begin{aligned}
J&= - \sum_{(u,q) \in \mathcal{S}} \ln x_{u, q}  \\
& \propto  - \sum_{\substack{(u,q) \in \mathcal{S} \\ l(q, e^{1})l(q, e^{2})...l(q, e^{N})l(q, i) = 1}} (\ln x_{u,e^{1}} + \ln x_{u,e^{2}} + ... + \ln x_{u,e^{N}} +\ln x_{u,i}). ~ 
\end{aligned}
\end{equation}

Since the explicit attributes are given in most e-commercial scenarios, the allocation $l(q, e)$ of items is usually known\footnote{We set this constraint for the reason that most recommender applications have explicit attributes information. Adding this constraint could make a great profit on performance. Without this constraint, the following optimization could still be done in the same way. Detail discussions are in Section \ref{sec:effect}.}.
We simplify our learning algorithm to only deal with the implicit attribute allocation $l(q, i)$.
Different from other parameters, the allocating function $l(\cdot)$ is non-differentiable. 
We thus take a bootstrap strategy \cite{li2016lightrnn} by iteratively conducting E-step and R-step. 
E-step mainly deals with estimating all latent attribute vectors, $e$ and $i$.
R-step aims to find a better implicit attribute allocation $l(q,i)$, since $l(q,e)$ is known (that is because the explicit attributes of an item are always known). 
The LearnDIR algorithm is shown in Algorithm \ref{alg}. 
Initially, we give each item a random implicit attribute. Then, we iteratively  conduct E-step and R-step until the convergence.

\begin{algorithm}[h]  
    \caption{LearnDIR}
	\label{alg}
    \begin{algorithmic}[1]
    	\STATE Randomly initialize the allocating function $l(\cdot)$;
    	\WHILE {the performance still improves in validation set} 
    		\REPEAT 
    		\STATE E-step: Train the specific recommender model with DIR to estimate latent vectors of explicit and implicit attributes and obtain $x_{u,e}$ and $x_{u,i}$;
        	    \UNTIL{the performance does not improve in validation set;} 	
    		\STATE R-step: Reallocate $l(\cdot)$ based on current $x_{u,e}$ and $x_{u,i}$. 
    	\ENDWHILE
    \end{algorithmic}
\end{algorithm}

\noindent \textbf{E-step: Estimating latent vectors.} 
This step assumes that the allocating function $l(\cdot)$ is given. With the fixed $l(\cdot)$, we train the specific model (e.g. MF, RNN) with DIR to learn the latent vectors of attributes.
The specific model estimates latent vectors of the explicit and implicit attributes, along with $x_{u,e}$ and $x_{u,i}$, which will be described in Section \ref{detail_model} in detail. 
Generally, the gradient of our loss function with respect to the model parameters $\Theta$ is,
\begin{equation} \label{eqh}
		\frac{\partial{J}}{\partial{\Theta}} \propto - \sum_{\substack{(u,q) \in \mathcal{S} \\ l(q, e^{1})l(q, e^{2})...l(q, e^{N})l(q, i) = 1}} (\frac{\partial{x_{u,e^{1}}}}{x_{u,e^{1}}\partial{\Theta}} + \frac{\partial{x_{u,e^{2}}}}{x_{u,e^{2}}\partial{\Theta}} + ... + \frac{\partial{x_{u,i}}}{x_{u,i}\partial{\Theta}}).
\end{equation}
We use stochastic gradient descent (SGD) to learn the parameter in E-step. In this case, for each user-item pair $(u, q) \in \mathcal{S}$, an update is performed.

\begin{equation} \label{eqh}
		\Theta \leftarrow  \Theta + \alpha (\frac{\partial{x_{u,e^{1}}}}{x_{u,e^{1}}\partial{\Theta}} + \frac{\partial{x_{u,e^{2}}}}{x_{u,e^{2}}\partial{\Theta}} + ... + \frac{\partial{x_{u,i}}}{x_{u,i}\partial{\Theta}}), ~
\end{equation}
where $\alpha$ is the learning rate of SGD. We stochastically update parameters by user-item pairs in the training set. The loss $J$ will converge after several epochs.

\noindent \textbf{R-step: Reallocating items to implicit attributes.}
This step fixes $x_{u,e}$ and $x_{u,a}$, which are estimated in E-step. 
Since $l(q,e)$ usually does not need to be learned, we remove the explicit part in  Equation (\ref{eqh_obj}), which leads to,
\begin{equation} \label{eqh}
\begin{aligned}
J & \propto  -\sum_{(u,q) \in \mathcal{S}, l(q,i)=1} \ln x_{u,i}\\
	& = -\sum_{(u,q) \in \mathcal{S}} \sum_{(q,i) \in \mathcal{Q} \times \mathcal{I}} l(q,i)\ln x_{u,i} \\
	&= -\sum_{(q,i) \in \mathcal{Q} \times \mathcal{I}} \left(l(q,i) \sum_{(u,q) \in \mathcal{S}} \ln x_{u,i} \right).~
\end{aligned}	
\end{equation}
Considering the constraints described in Equations (\ref{constraint1}) and (\ref{constraint2}), the complete optimization problem can be summarized as,
\begin{equation} \label{eqh}
\begin{aligned}
		\mathop{\max}_{l(\cdot)} \quad &\sum_{(q, a) \in \mathcal{Q}_{c} \times \mathcal{A}} \left(l(q, a) \sum_{(u,q) \in \mathcal{S}} \ln x_{u,a}\right),~ \\
		\text{s.t.} \quad & \sum_{q \in \mathcal{Q}_{c}} l(q, a) \leq 1, \forall a \in \mathcal{A},~\\
		\quad & \sum_{a} l(q, a) = 1, \forall q \in \mathcal{Q}_{e},~ \\
		\quad & l(q, a)  =\{ 0, 1\}, ~
\end{aligned}
\end{equation}
where $\mathcal{Q}_{e}$ denotes the set of items having the same combination of explicit attributes $e^{1}, e^{2}, ..., e^{N}$. 

For each $\mathcal{Q}_{e}$, 
we use the OR-tool\footnote{https://developers.google.com/optimization/; we use the solver of minimizing cost flow.} to solve this optimization as finding a maximum matching between the item set $\mathcal{Q}_{e}$ and the implicit attribute set $\mathcal{I}$. 

\subsection{Model discussion}
The time complexity of R-step is $O(I^{3} \log I)$ \cite{ahuja2017network}, where $I$ is the number of latent vectors of implicit attributes. The training procedure could be conducted off-line, where the time consumption is acceptable. 
Both E-step and R-step share the same objective function. The only difference is E-step optimizes continuous variables, and R-step optimizes discrete variables.  Since both step has the same objective function, they will promote each other. Similar strategies have been used in different application. \cite{li2016lightrnn, mcauley2013amateurs} We do experiments in Section \ref{sec:convergence} for further discussion.

DIR framework largely cuts down the parameters of item representation and does not bring much time complexity during inference. However, the drawback of DIR based methods is the time complexity in training process. The training process is much longer than original methods. The main problem comes from the difficulty of allocating items into suitable attributes (i.e., MCMF). In real-world application, we could give a good initialization of allocation according to prior knowledge or previous training results. 
Besides, the inference time of DIR is similar to item based representation models, which makes DIR workable.

DIR framework could alleviate the cold-start problem to a certain extent without adding other content information. In common collaborative methods, the cold-start problem partly comes from too sparse collaborative information of new items. It is difficult for new items to have accurate representations, since there are few collaborative information for optimization. 
 However, under the framework of  DIR, all items are represented at attribute level, which means items share attribute vectors with others, so that a new item could share well-learnt attributes vectors of other old items.
 We can learn how to best represent or allocate new items based on the already well-learned attributes in R-step, 
which only need a few collaborative records.

\subsection{Learning Models with DIR} \label{detail_model}
 \noindent DIR  is a model-agnostic attributed based item representation framework. All the collaborative filtering based methods who represent items as vectors and calculate $x_{u,q}$ for recommendation could be used under DIR.  
In this subsection, we describe two typical recommender models, MF and RNN, and show how we can learn these two models with DIR. 
Both MF and RNN try to predict the possibility of a user purchasing an item. For MF, the prediction is a score $x_{u,q}$ per user-item pair $(u,q)$. For RNN, the prediction is $x_{u,q,t}$ per user-item-time tuple $(u,q,t)$. 
Instead of modeling an item by a latent vector in original MF and RNN, we provide a latent vector for each attribute and model items by a combination of attribute vectors in the DIR framework.

\noindent \textbf{MF with DIR.}
To predict the rating score $x_{u,q}$ of user $u$ on item $q$, MF represents both users and items in a joint latent vector space, such that the user-item interaction $x_{u,q}$ is modeled as the inner product in that space \cite{koren2009matrix}. 

In DIR, MF could model the preference of user u for explicit and implicit attribute by dot product as \begin{equation} \label{eqh_uac}
	 x_{u,e} = \langle \mathbf{u},  \mathbf{e} \rangle \quad
	 x_{u,i} = \langle \mathbf{u},  \mathbf{i} \rangle.
\end{equation}
According to Equations (\ref{eqh_obj}) and (\ref{eqh_uac}), the objective function in DIR-MF is expressed as, 
\begin{equation}
J =  - \sum_{\substack{(u,q) \in \mathcal{S} \\ l(q, e^{1})l(q, e^{2})...l(q, e^{N})l(q, i) = 1}} ln \langle \mathbf{u}, \mathbf{e^{1}} \rangle + ln \langle \mathbf{u}, \mathbf{e^{2}} \rangle + ... + ln \langle \mathbf{u}, \mathbf{e^{N}} \rangle + ln \langle \mathbf{u}, \mathbf{i} \rangle. ~
\end{equation}

\noindent \textbf{RNN with DIR.}
RNN can capture the sequential structure of the user historical behavior. 
To describe RNN for recommendation, we formulate $q_{t}^{u}$ to denote the item which is bought by $u$ at time step $t$. The behavior set $\mathcal{S}^{T}$ is formed by the triple $(u, q, t)$ which means $u$ purchases $q$ at time step $t$. 
 
In the RNN recommender model \cite{yu2016dynamic,liu2016context,cui2019hierarchical}, the input sequence of RNN is the item representation $\mathbf{q}_{t}^{u}$  of the item $q_{t}^{u}$.  
At time step $t$, we calculate the hidden layer $\mathbf{u}_{t}$ by the input $\mathbf{q}_{t}^{u}$ and the hidden layer $\mathbf{u}_{t-1}$ at the previous time step $(t-1)$.
The hidden layer $\mathbf{u}_{t}$ can be treated as the dynamic representation of the user $u$. 
The following equation denotes a unit of RNN,
\begin{equation} \label{eqh}
	\mathbf{u}_{t} = f (\mathbf{W} \cdot \mathbf{q}_{t}^{u} + \mathbf{V} \cdot \mathbf{u}_{t - 1}),~
\end{equation}
where $\mathbf{W}$, $\mathbf{V}$ are transition parameter matrices and $f(\cdot)$ is the activation function (e.g., the sigmoid function $f(x) = \frac{1}{1 + e^{-x}}$).	

In DIR-RNN, we concatenate the attribute vectors to represent the corresponding item $\mathbf{q}_{t}^{u} = [\mathbf{e^{1}}; \mathbf{e^{2}}; ... ; \mathbf{e^{N}}; \mathbf{i}]$. 
That is to say,
\begin{equation} \label{eqh}
	\mathbf{u}_{t} = f (\mathbf{W} \cdot [\mathbf{e^{1}}; \mathbf{e^{2}}; ... ; \mathbf{e^{N}}; \mathbf{i}] + \mathbf{V} \cdot \mathbf{u}_{t - 1}),  ~
\end{equation}
where the item $q_{t}^{u}$ belongs to the explicit attribute $e^{1}, e^{2}, ..., e^{N}$ and implicit attribute $i$. 

At each time step, the preference of user $u$ for explicit and implicit attributes could be determined by dot product,
\begin{equation} \label{eqh_xuat_xuct}
	x_{u,e,t} = \langle \mathbf{u}_{t}, \mathbf{e}\rangle \quad
	x_{u,i,t} = \langle \mathbf{u}_{t}, \mathbf{i}\rangle.
\end{equation}

Combining (\ref{eqh_xuat_xuct}) with (\ref{eqh_obj}), the objective function of DIR-RNN is expressed as,
\begin{equation} \label{eqh}
J = - \sum_{\substack{(u, q, t) \in \mathcal{S}^{T} \\ l(q, e^{1})l(q, e^{2})...l(q, e^{N})l(q, i) = 1}}
ln \langle \mathbf{u_{t}}, \mathbf{e^{1}} \rangle + ln \langle \mathbf{u_{t}}, \mathbf{e^{2}} \rangle + ... + ln \langle \mathbf{u_{t}}, \mathbf{e^{N}} \rangle + ln \langle \mathbf{u_{t}}, \mathbf{i} \rangle.~
\end{equation}

\section{Experiments}

 \noindent The main contribution of this work is to develop a new representation method DIR for recommendation. We aim to answer the following research questions via experiments.
 \begin{itemize}
    \item[] \textbf{RQ1} \quad Does the method under the DIR framework perform better than other methods?
 	 \item[] \textbf{RQ2} \quad Does the method under the DIR framework perform better than other methods in cold-start situation?
 	 \item[] \textbf{RQ3} \quad Does the improvement of performance come from the disentangled representation at the attribute level?
 	\item[] \textbf{RQ4} \quad How does the learned attribute representation guide the real-world application?
 	\item[] \textbf{RQ5} \quad How do different parts of DIR affect the performance and how does DIR work?
 \end{itemize}
 Next, we first describe the experimental settings. We then report results by answering the above research questions in turn.

\begin{table}[h]
\centering
\caption{Statistics of datasets}
    \begin{tabular}{cccccc}
    \toprule
Dataset & \# Users & \# Items & \# Category & \# Feedback & \% Cold-start \\
    \midrule
     Clothing & 20196 & 22280 & 323 & 145932 & 34.25\\ 
    \midrule
    Electronics & 15944 & 28519 & 281 & 117485 & 22.25\\    
    \bottomrule
    \end{tabular}%
\label{tab:datasetdiscribe}%
\end{table}%

\subsection{Experimental Settings}
\noindent \textbf{Datasets:} We use the amazon dataset\footnote{http://jmcauley.ucsd.edu/data/amazon/links.html}~\cite{mcauley2015image,he2016ups} to assess the performance of our proposed method. We choose the ``Clothing Shoes \& Jewelry'' and the ``Electronics'' datasets, which are named as \textbf{Clothing} and \textbf{Electronics} for short. 
Following the previous work \cite{sun2017exploiting}, 
we uniformly sample the datasets to balance their sizes for the cross-dataset comparison.
Following the work of \cite{liu2017deepstyle}, we remove users with less than 5 purchases or more than 100 purchases. 
The statistics of datasets are listed in Table \ref{tab:datasetdiscribe}.
Items which appear less than 5 times in the training data are called cold-start items.
The percent of cold-start items is $34.25\%$ of overall testing set in Clothing, while $22.25\%$ in Electronics.
We evaluate all methods with two settings in the following sections: Warm-start and Cold-start. The warm-start setting focuses on the overall ranking performance using the whole test set, while the cold-start setting only uses cold-start items in the test set.

\noindent \textbf{Parameter Settings:} 
Since category is the only completely given attribute from these two datasets, we only consider one explicit attribute (i.e., category) in our experiment. 
In both DIR-MF and DIR-RNN, we initially set the learning rate as 1.0 and then decrease it by a ratio of 2 after ten epochs. 
In our experiment, the number of implicit attributes is 1584 in the Clothing dataset, and 2958 in the Electronics dataset. 
The number of latent vectors of implicit attributes is equal to the number of items in the largest category, which is the least number of implicit attributes to guarantee  that every item has a distinctive combination (detail discussion in Section \ref{sec:implicit_num}).
In order to have a better category representation, we embed categories in a hierarchical way, similar to Sherlock \cite{he2016sherlock}, which has been proved effective and efficient.

\noindent \textbf{Evaluation Protocol:} Similar to several previous works of recommendations with implicit feedback \cite{rendle2009bpr,he2016sherlock}, we also use the leave-one-out protocol for evaluation. For each user $u$, we use the last purchase for testing and the others for training. 
We apply the average AUC (Area Under the ROC Curve) metric to evaluate the performance of  all methods in terms of personalized ranking, 
\begin{equation} \label{eqh}
		AUC= \frac{1}{\lvert \mathcal{U}\lvert} \sum_{u \in \mathcal{U}}  \frac{1}{\lvert \mathcal{Q} \backslash \mathcal{Q}_{u} \lvert} \sum_{r \in  \mathcal{Q} \backslash \mathcal{Q}_{u}} \delta (x_{u,p}>x_{u,n}), ~
\end{equation}
where $p$ is the leave-out item that the user $u$ purchases, $n$ is a random item that the user $u$ hasn't purchased, $\delta (\cdot)$ is the indicator function which equals to one when the condition is met, and zero otherwise.

\noindent \textbf{Compared methods:} we compare the proposed model with the following methods:
\begin{itemize}
\item \textbf{MF} \cite{rendle2009bpr} is required as BPR-MF which is known as a classical MF method for implicit feedback.
\item \textbf{FPMC} \cite{rendle2010factorizing} combines Markov chain and MF for sequential recommendation.
\item \textbf{RNN} \cite{yu2016dynamic} here denotes the LSTM-based RNN model, which alleviates the long-term dependencies of basic RNN.    
\item \textbf{Sherlock} \cite{he2016sherlock} uses both hierarchical category information and visual information to enhance the item embeddings.
\item \textbf{HieVH} \cite{sun2017exploiting} models the item relationship using hierarchical category information.
\end{itemize}


\begin{table*}[htbp]
\small
\centering
\caption{Recommendation performance on the Clothing and Electronics datasets evaluated by AUC. Here, the number of parameters includes all the parameters a model needs to learn, where M means one million. Time records the the inference time that each method needs for entire test set.}
\vspace{1mm}
    \begin{tabular}{c|c|ccccc|cc}
    \toprule
    Dataset & Setting & MF & FPMC & RNN  & Sherlock & HieVH & DIR-MF & DIR-RNN \\
    \midrule
\multirow{3}[0]{*}{Clothing} & warm-start & 
0.5842 & 0.6766 & 0.6784  & 0.7065 & 0.7401 & 0.6549 & \textbf{0.7569} \\

& \# parameter (M) & 1.3003 & 6.1533 & 1.1342 & 12.2634 & 496.8363 & 1.1373 & \textbf{0.2886} \\
& time (s) & 0.1134 & 0.2932 & 120.3526 & 0.2767 & 1.1592 & 0.2738 & 178.1951 \\

    \midrule
\multirow{3}[0]{*}{Electronics} & warm-start &
 0.7061 & 0.7964 & 0.7517  &0.7492 & 0.8217& 0.7585 & \textbf{0.8548}  \\
& \# parameter (M)& 
1.3339 & 5.0751 & 1.4462 & 7.7214 & 816.0235 & 0.7594  & \textbf{0.2720} \\
&  time (s) & 0.0806 & 0.1835& 45.8593 & 0.2418 & 0.8251 & 0.2379 & 59.1175\\

    \bottomrule
    \end{tabular}%
\label{tab:compare}%
\end{table*}%

 \begin{figure}[htbp]
\centering

\subfigure[Clothing]{
\begin{minipage}[b]{0.8\textwidth}
\label{fig:size_auccloth} 
\includegraphics[width=1\textwidth]{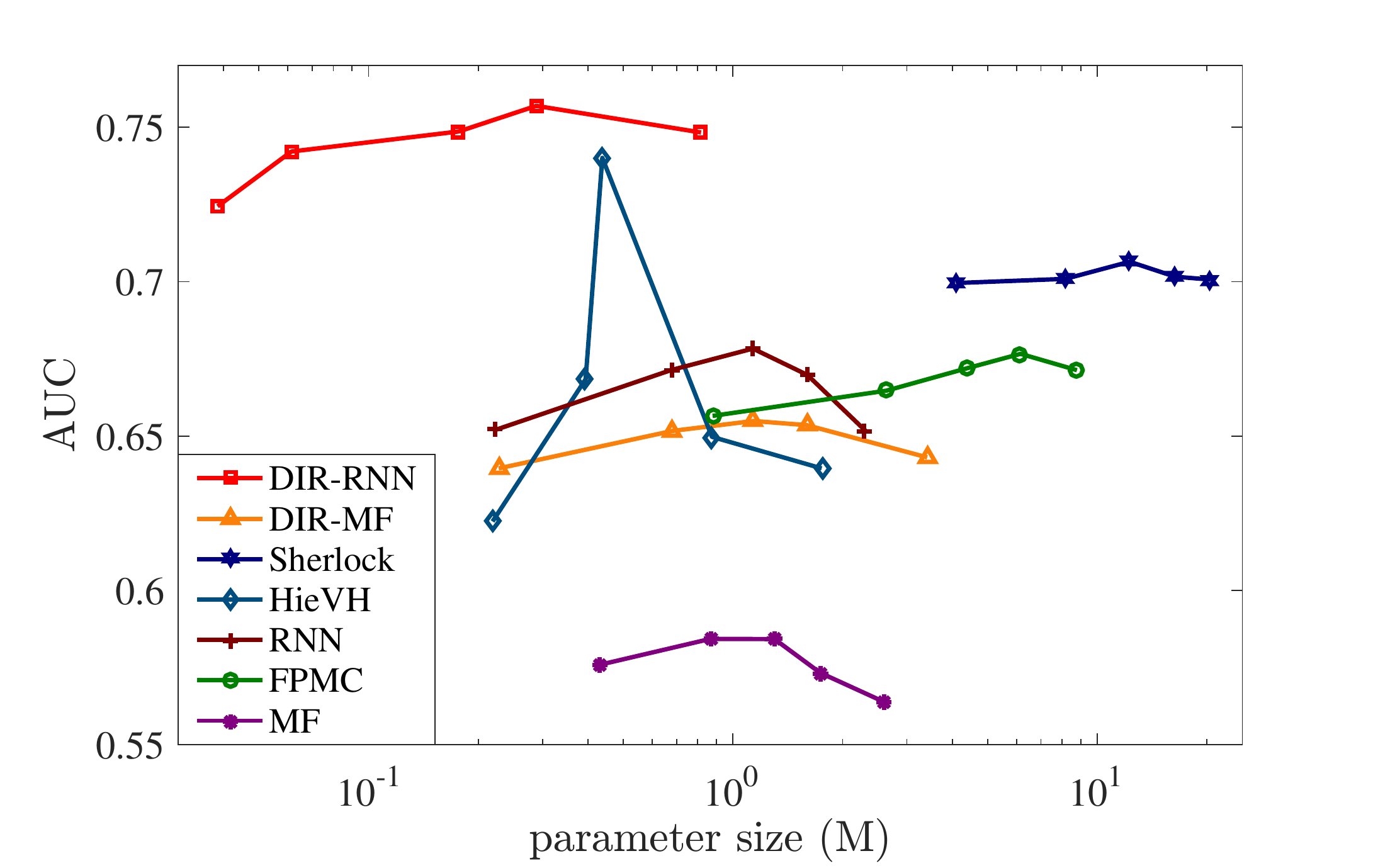}
\end{minipage}%
}%

\subfigure[Electronics]{
\begin{minipage}[b]{0.8\textwidth}
\label{fig:size_aucelect} 
\includegraphics[width=1\textwidth]{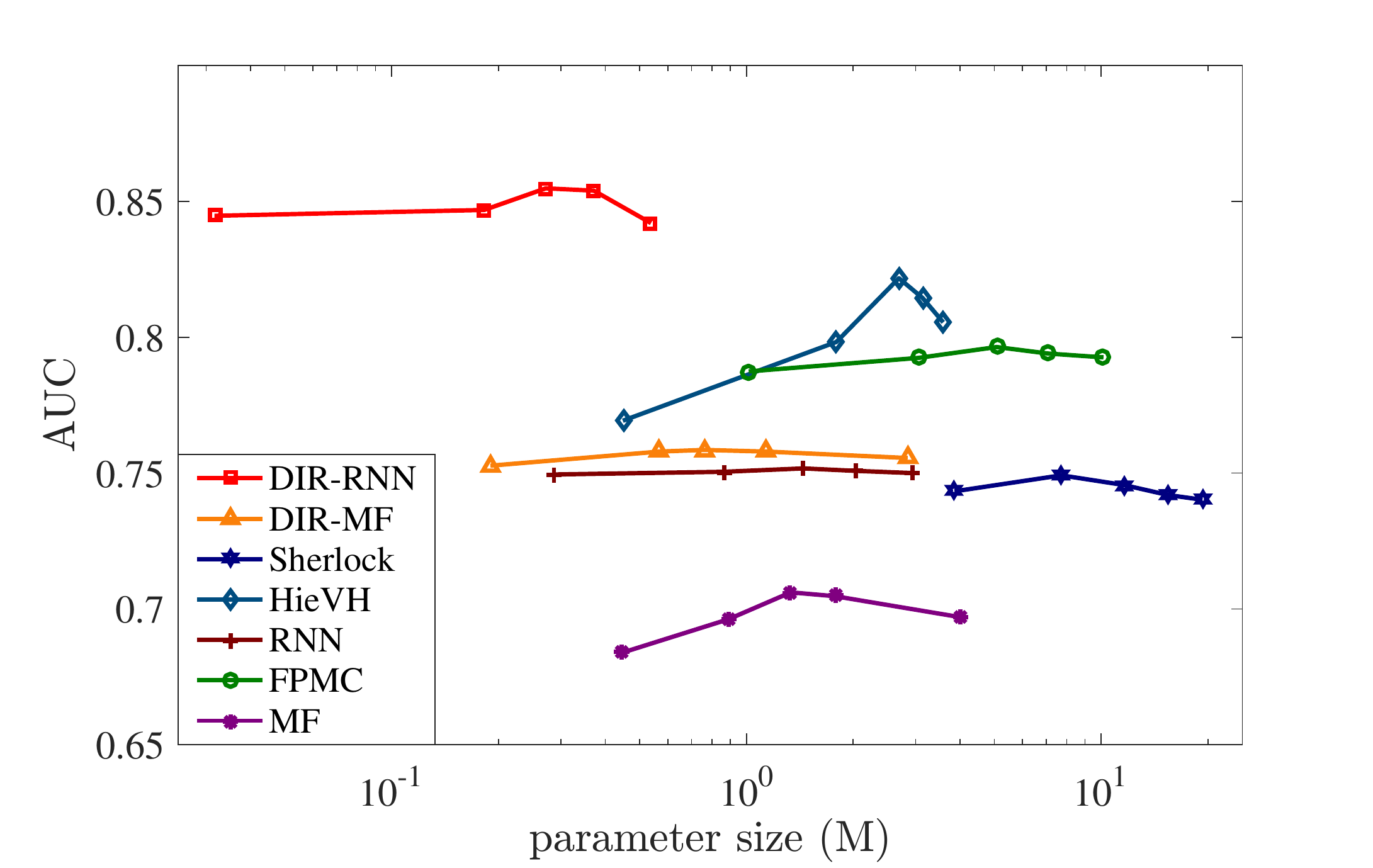}
\end{minipage}%
}%

\caption{AUC performance of different models with different sizes of parameters}
\label{fig:size_auc}
\end{figure}

\
\subsection{Performance Comparison (RQ1)}
\noindent  
We list the best performance, parameter size and the corresponding inference time of each method in Table \ref{tab:compare}. 
First of all, 
we can see that DIR-MF and DIR-RNN significantly improve the performance of both MF and RNN, which illustrates that DIR can well describe item properties for different recommendation models. 
Among the compared models, 
FPMC and RNN are both sequential recommendation models, and perform  well in two datasets. 
Although HieVH achieves a relatively higher AUC in both datasets, it needs a complicated pre-training for item co-occurrence matrix consuming too much parameter space, which is difficult to use in real-world applications.
Sherlock is highly dependent on the representativeness of visual information, so it performs well only in the Clothing dataset, and relatively poor in the Electronic dataset where visual information has difficulty in revealing attributes. 
Different from these methods, DIR-based models are not dependent on visual information. They perform stably in different datasets.
Especially, DIR-RNN outperforms all other methods in terms of the prediction quality, 
which illustrates the ability of DIR.
In term of evaluation time, DIR-MF consumes similar time with  other MF based methods. DIR-RNN is also similar to RNN. The result proves that DIR does not need much additional evaluating time

Figure \ref{fig:size_auc} shows the AUC of all models on two datasets with different parameter sizes. The traditional methods tend to obey the rule that better performance relies on more parameters. While our DIR-based methods, especially DIR-RNN (on the left-top corner), use relatively fewer parameters and achieve the best performance among other methods.
Although DIR-MF does not extremely outperform other methods, it obtains a great improvement compared with BPR-MF.

\begin{table*}[htbp]
\centering
\caption{Recommendation performance on the Clothing and Electronics datasets evaluated by AUC in cold-start situation. }
    \begin{tabular}{c|ccccc|cc}
    \toprule
    Dataset & MF & FPMC & RNN  & Sherlock & HieVH & DIR-MF & DIR-RNN \\
    \midrule
Clothing &
0.5315 & 0.5574 & 0.5182  & 0.6498 & - & 0.5459 & \textbf{0.6621} \\

    \midrule
Electronics & 
 0.5126 & 0.5868 & 0.5897 & 0.5823 & 0.5325 & 0.5332 & \textbf{0.6491} \\
    \bottomrule
    \end{tabular}%
\label{tab:cold_start}%
\end{table*}%

\subsection{Performance Comparison in Cold-start Situation (RQ2)}
 \noindent The sparser training data is, the more difficult learning representation of items will be. 
In order to test the representation ability of DIR, we compare the methods in the cold-start situation. 
Table \ref{tab:cold_start} shows  AUC results of all methods under the cold-start situation of both Clothing and Electronic datasets.
DIR-RNN achieves the best performance in the cold-start setting of two datasets and obtains an average improvement of 6.68\% compared with Sherlock and 5.23\% compared with FPMC. In Clothing, content information describes items well, so Sherlock gets a good performance. In Electronics, 
the sequential information plays an important role for recommendation, which is the reason for the exceptionally good performance of sequential models, i.e., FPMC, RNN and DIR-RNN.
Our DIR-RNN performs better in both datasets, which means DIR can better describe the properties of items.  

\begin{figure}[htbp]
\centering

\includegraphics[width=0.8\textwidth]{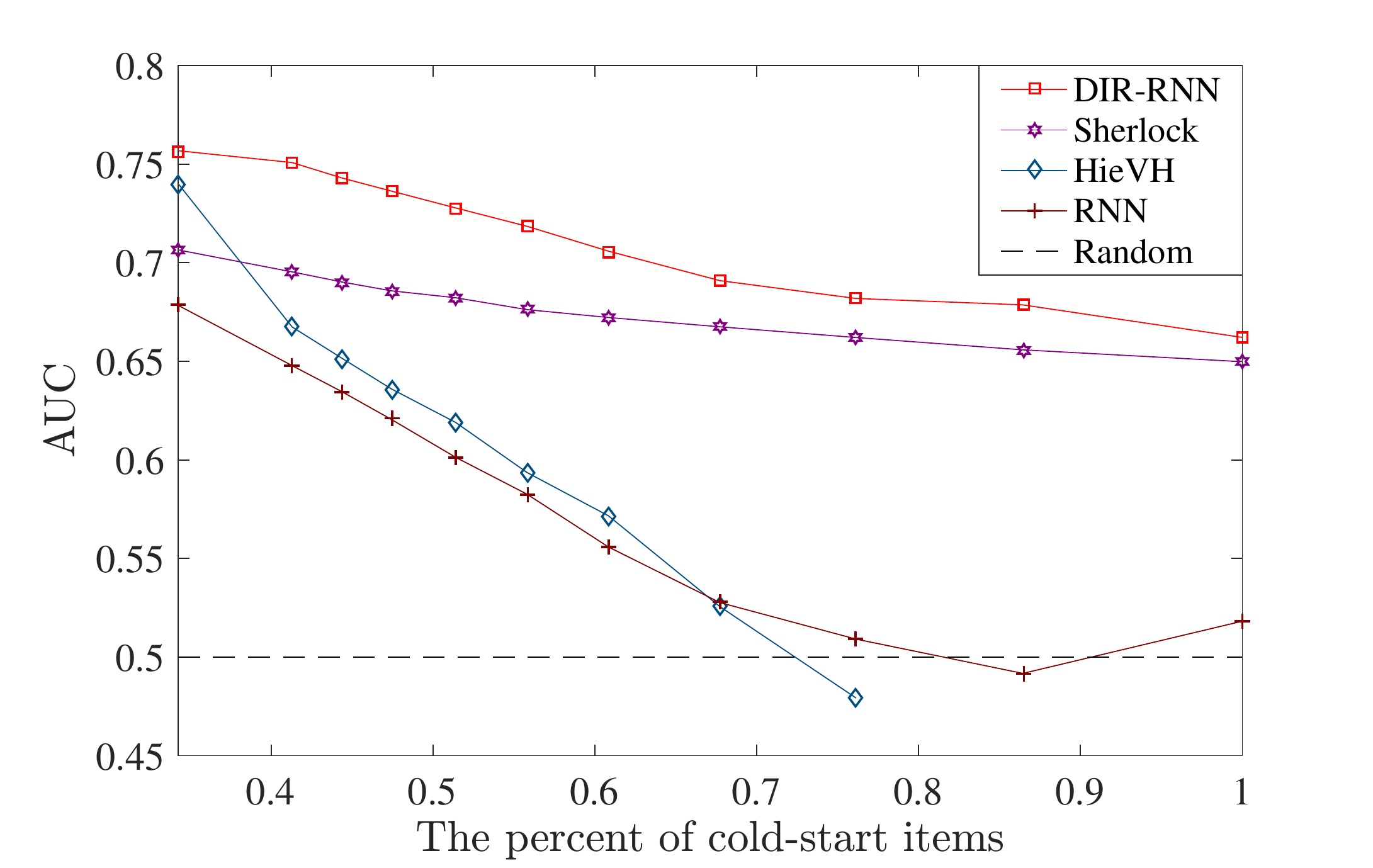}
\caption{The performance of different methods with different degrees of cold-start situation. X-coordinate represents the percent of cold-start items in the testing set.}
\label{fig:cold_start}
\end{figure}

In Figure \ref{fig:cold_start}, we look into the performance of several methods with different degrees of cold-start situation. As can be seen in the figure, the AUC results decrease as the percent of infrequent items increasing.  
HieVH performs poor in the cold-start situation. When the infrequent percent improves to 0.8, the model completely fails. 
Only DIR-RNN and Sherlock could maintain a stable performance in the cold-start situation. Sherlock takes advantage of visual information to achieve better representations of items. So it could also learn the infrequent items in the training set well. DIR-RNN even performs better than Sherlock, especially with 0.4 to 0.6 percent of infrequent items. 
Overall, the results show that DIR could learn the item representations well even if there is inadequate supervised information. 
The key reason why DIR could have a nice performance in cold start situation is that DIR makes items share attribute vectors with others, so that a new item could share well-learnt attributes vectors of other old items.
In cold start situation, the user-item pairs are sparse but the user-category pair or user-implicit attribute pairs are always dense, making the representation easy to learn. In another words, for a new item who only has a few records,
we only need to learn how to best represent or allocate new items based on the already well-learned attributes.
%
%
%

\begin{table*}[htbp]
\centering
\caption{The comparison of attribute disentangled representation model (DIR-*) and attribute augmented representation model (Augmented-*). The parameter sizes of the corresponding methods are listed in the table, where M means one million. Time records the the inference time that each method needs for test set.}
    \begin{tabular}{l|ccc|ccc}
    \toprule
   Datasets & \multicolumn{3}{c|}{Clothing} & \multicolumn{3}{c}{Electronics} \\
\midrule
   Setting & AUC & \# param (M) & time (s) & AUC & \# param (M) & time (s) \\
    \midrule
Augmented-MF  & 
0.6256  & 1.9419 & 0.2745 & 0.7128 & 1.8206 & 0.2467 \\
Augmented-RNN & 0.7074  & 1.1504 & 221.0825 & 0.7623 & 1.4602 & 70.4628 \\
    \midrule
DIR-MF($e^{-}$)  & 
0.6524  & 2.1364 & 0.2485  & 0.7182 & 1.1397 & 0.1865  \\

DIR-RNN($e^{-}$) & 0.6631  & 0.1504 & 139.5490 & 0.8185 & 0.0827 & 47.0039  \\
    \midrule
DIR-MF  &
 0.6549  & 1.1373 & 0.2738  & 0.7585 & 0.7594 & 0.2379 \\
DIR-RNN  & 
\textbf{0.7569}  & \textbf{0.2886} & 178.1951  & \textbf{0.8548}  & \textbf{0.2720} & 59.1175 \\
    \bottomrule
    \end{tabular}%
\label{tab:concate}%
\end{table*}%

\subsection{Effectiveness of Disentangled Item Representation (RQ3)}\label{sec:effect}
In this subsection, we further discuss the advantages of DIR. We make experiments in two aspects. Firstly, we compare DIR with with other kinds of augmented item representations. They use the same attribute features as DIR, this comparison could reveal the effectiveness of DIR. Secondly, we remove the explicit attribute information, and allocate the every items into 2 dimension matrix, which could investigate the advantages of explicit information. 

\noindent \textbf{Augmented-MF} simply concatenates the vectors of attribute information with the latent vectors of items. This augmented latent vectors are used as item representations in MF.

\noindent \textbf{Augmented-RNN} is similarly designed with Augmented-MF, and the same augmented vectors are used for the RNN recommendation model.

\noindent \textbf{DIR-MF($e^{-}$)} 
Similar model as DIR-MF, while we do not use any explicit information. We allocate every items to 2 attributes. Both the attributes of each items are randomly initialized and re-allocated by LearnDIR algorithm. The numbers of vectors in both dimension are equal to the upper rounding of the root of item numbers, to ensure each item has unique attributes combination.

\noindent \textbf{DIR-RNN($e^{-}$)} DIR-RNN with the same setting as  DIR-MF($e^{-}$).
 \noindent
 
Among Augmented-* and DIR-*, the attribute information is processed in the same way, i.e., each attribute is represented by a vector of $d$ dimensions. For fair comparison, Augmented-* shares the same settings with its corresponding DIR-*. For convenience, we only consider the attribute information of categories. 
DIR-*($e^{-}$) is similar with the setting of DIR-*. Instead of   using explicit attribute, DIR-*($e^{-}$) disentangles  item representation into two implicit attributes.  
The result of each method is given in the Table \ref{tab:concate}. DIR-MF (DIR-RNN) achieves great improvement compared with Augmented-MF (Augmented-RNN), which proves the superiority of DIR over other augmented item representations.    
Models under the framework of DIR use less parameters compared with the other models, since DIR lets different items share the same representation of attributes. 
DIR-*($e^{-}$) also gets well performances in both two datasets, even better than Augmented-* models. That is to say, items could find their attributes during LearnDIR.
To sum up, DIR can achieve better performance compared with traditional embedding method with fewer parameters.

\subsection{The Guidance for Real-World Applications (RQ4)}
 \noindent
In this section, we will give a general display of DIR's guidance for real-world applications. 

Firstly, the learned attribute representations of DIR can better reflect different properties of items.
Secondly, since we represent items at the attribute level, the user preference for items can be described at the attribute level, which can provide explanation for users in recommender systems. 

\begin{figure}[htbp]
\centering
\includegraphics[width=0.6\textwidth]{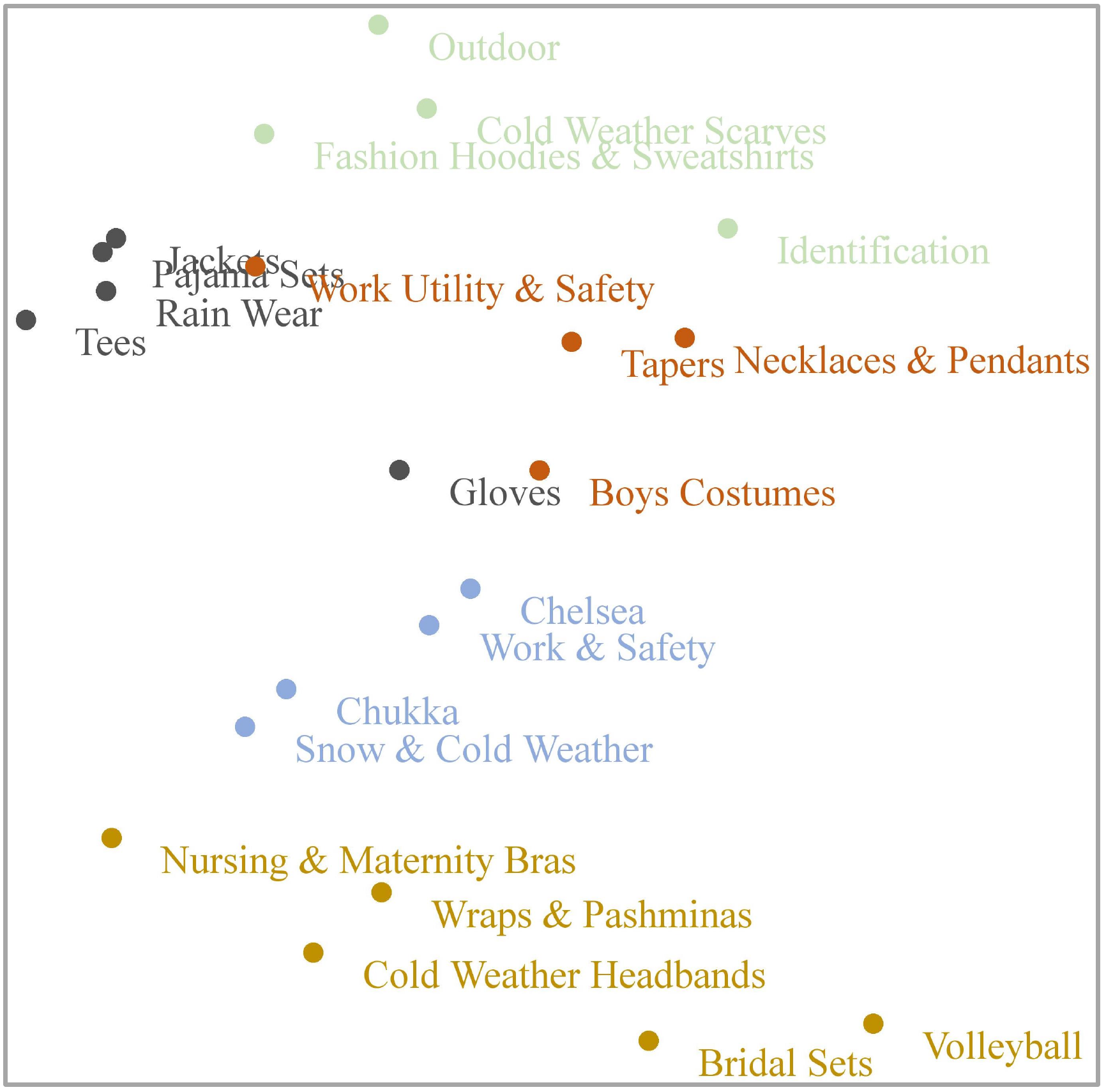}
\caption{The 2-D PCA visualization of category vectors (better viewed in color). The categories with the same color belong to the same father category. Similar categories are distributed in similar positions in the vector space.}
\label{fig:cate}
\end{figure}

\subsubsection{Visualization of Different Attribute Representations} \label{sect:vis}
 \noindent
First, we depict the category vectors of DIR. In Figure \ref{fig:cate}, we project category representations to 2-D vectors by principal component analysis (PCA). Categories with the same color are from the same father category. For example, orange is ``Novelty Costumes \& More'', yellow is ``Women'', black is ``Boys'', green is ``Men'' and blue is ``Boot''.
It is obvious that categories with the same color cluster, which indicates that similar categories learn similar embedding vectors. 
In that case, category representations in our model are well-learnt.

\begin{figure}[htbp]
\centering

\includegraphics[width=0.7\textwidth]{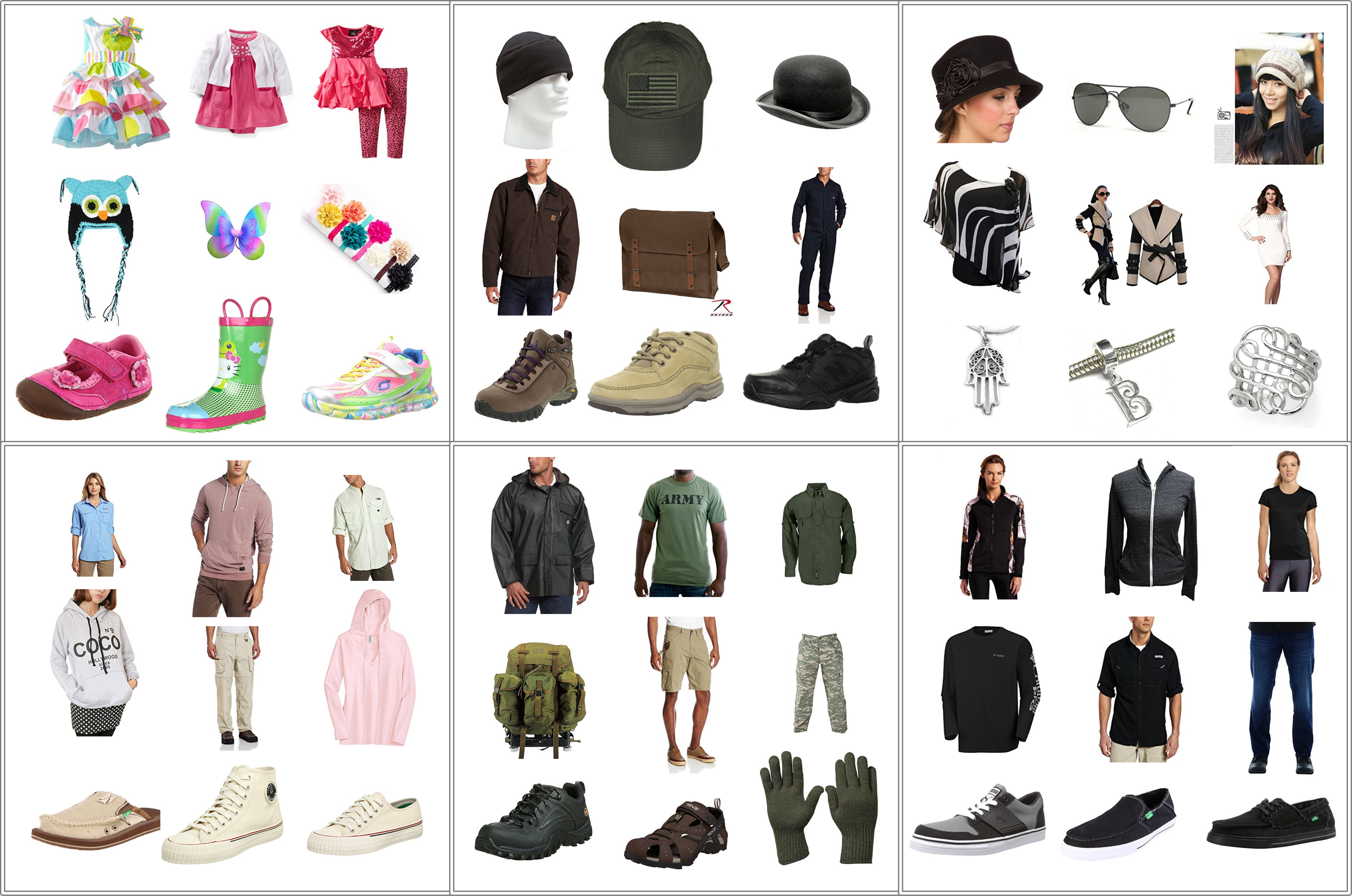}
\caption{Some clustering results of clothing items learnt by DIR-RNN on the Clothing dataset. Items in one box are with the same implicit attribute and belong to different categories.}
\label{fig:visualclothing}

\end{figure}

Then, we cluster items according to implicit attribute representations and visualize them in a table as shown in Figure \ref{fig:visualclothing}. Items in the same box belong to the same cluster. As can be seen, the learnt implicit attribute of clothing can reflect style regardless of category information.
From left to right and from top to bottom, the implicit attributes tend to be cute-like, casual-like, working uniform, out-door actives, fashionable-like and sports-like.  
This proves the effectiveness of the learnt attribute representations.

\begin{figure}[htbp]
\centering
\includegraphics[width=0.7\textwidth]{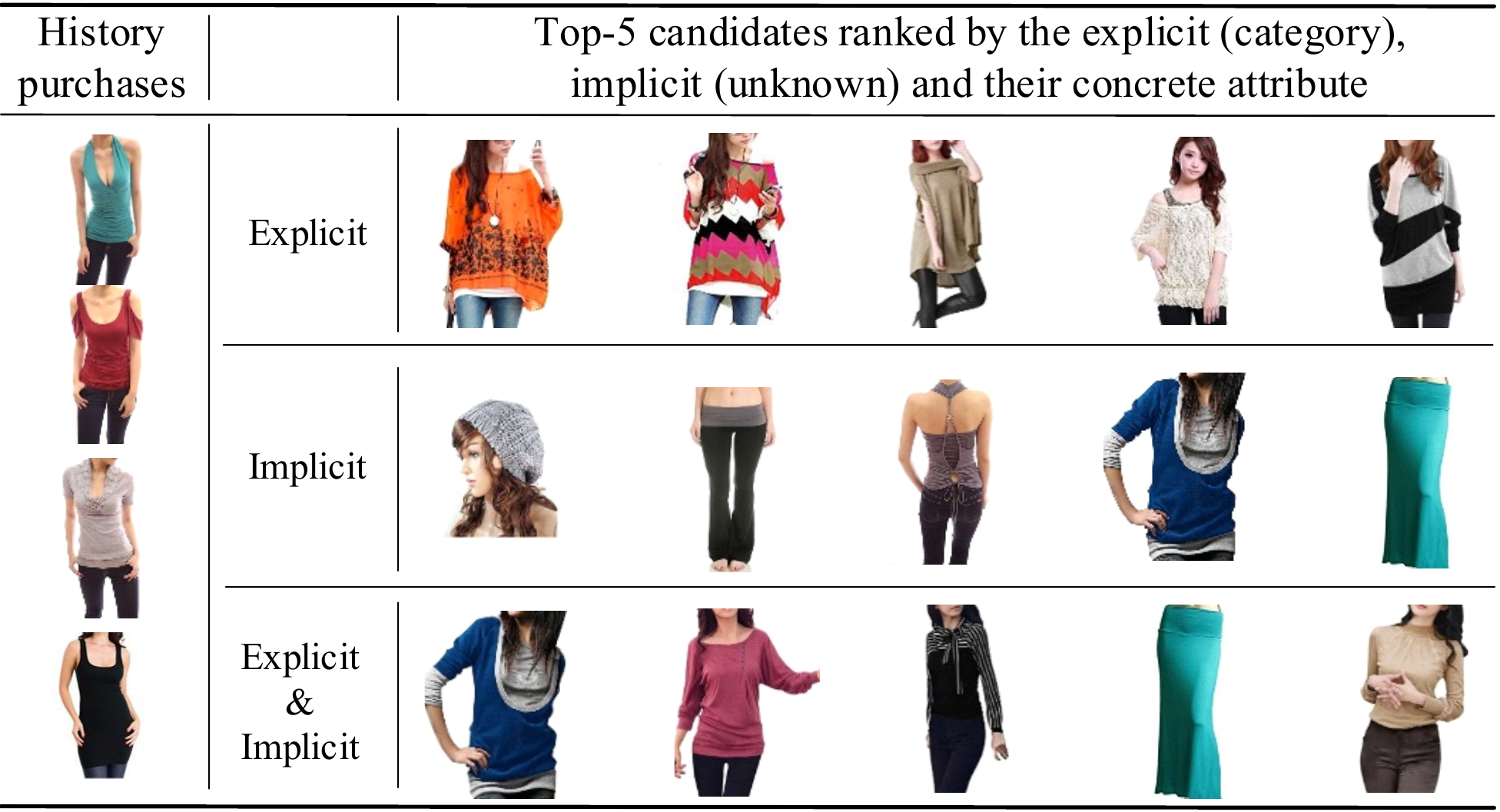}
\caption{A case study of the  attribute-level preference  on a random user. Historical purchases of the user are on the left; the recommendation items learned by our model are on the right. Each row of items is top-5 items ranked by the explicit (category), implicit (unknown) and their integrated attribute. The top row is ranked by the category attribute. The second row is ranked by the unknown attribute which is mainly about style. The items in the third row are the candidates considering both.}
\label{fig:casestudy}
\end{figure}

\subsubsection{Preference at the Attribute Level} \label{sect:case}
 \noindent  
Traditional collaborative methods can only predict the users' preference at the item level, while methods under the framework  of DIR can predict the preference at the attribute level.
Here we show the users' preference for the category attribute and implicit attributes of items under the framework of DIR.
We randomly look into a user who has bought some clothing, shown on the left side of Figure \ref{fig:casestudy}. 
We use our proposed DIR-RNN to find the top-100 items as the candidate set for recommendation, then calculate the preference for category attribute, implicit attribute and also the item (product of the two above preferences). 
The top-5 ranking results are listed in the right side of Figure \ref{fig:casestudy}. From top to bottom, they indicate preference according to category attribute, implicit attribute, and the integration of them. 
The user's  interests is low-chest, uniform color, tight  women's blouses. Among this attributes, ``women's blouses'' is the explicit attributes (i.e., category information), while ``low-chest'', ``uniform color'' and ``tight'' are implicit attributes in the dataset.
In the top row, the model recommends items according to the woman's preference for category, i.e., the women's blouses that the user often bought. 
In the second row, the top-5 items are of the same implicit attribute with the user's historical purchases, which capture the user's preference for style (as described in Sect \ref{sect:vis}), i.e., low-chest, tights and uniform color. 
Items listed in bottom meet both the user's preference for category and style. i.e., tight and solid color blouses, which proves the precise and diverse recommendation quality of DIR.

\begin{figure}[htbp]
\centering

\subfigure[Clothing]{
\begin{minipage}[b]{0.4\textwidth}
\label{fig:cloth1} 
\includegraphics[width=1\textwidth]{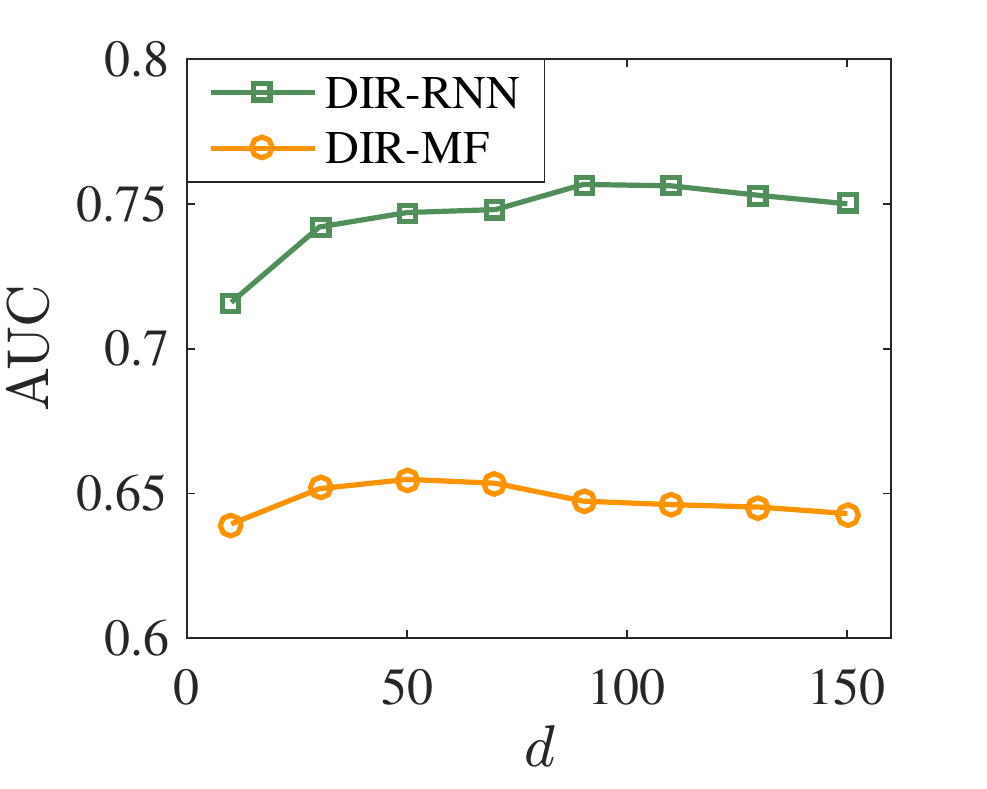}

\end{minipage}%
}%
\subfigure[Electronics]{
\begin{minipage}[b]{0.4\textwidth}
\label{fig:cloth2} 
\includegraphics[width=1\textwidth]{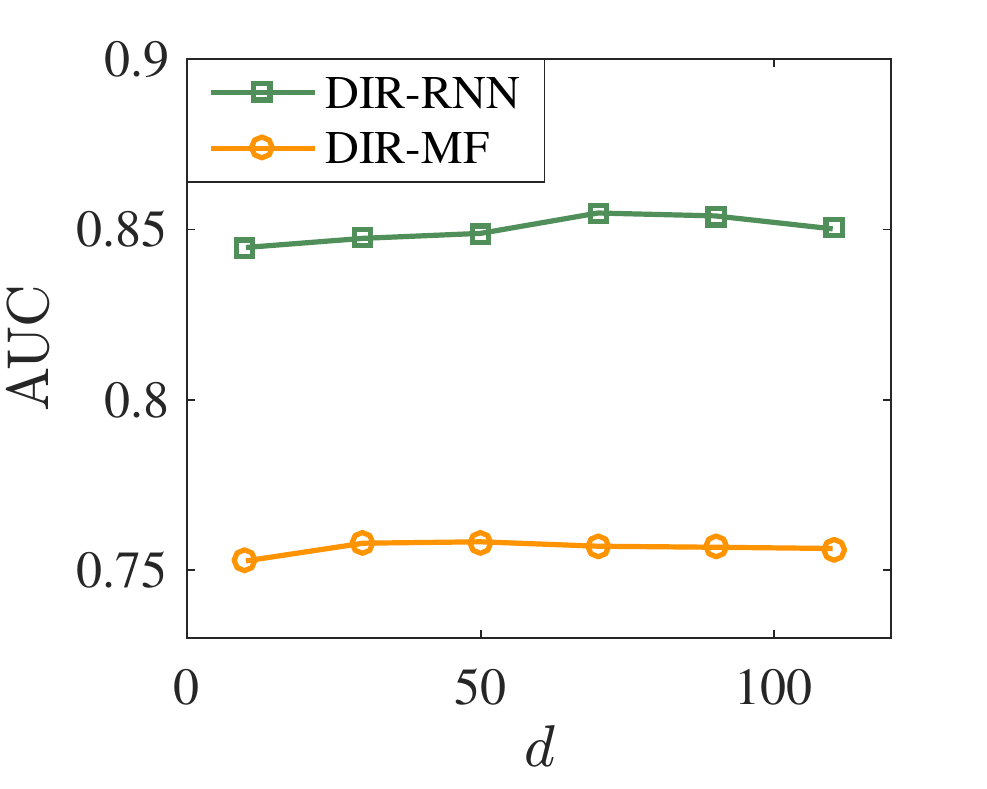}
\end{minipage}%
}%
\caption{AUC performance with different dimensionalities $d$ on two datasets (Clothing is on the left, Electronics on the right).}
\label{fig:performance}
\end{figure}

\subsection{Model Discussion (RQ5)}
 \noindent
This section mainly discusses the feasibility of our model. First, we discuss the how hyper parameters affect the performance. Then, we show the convergence of our methods during training.
\subsubsection{Impact of Embedding Dimensionalities}
 \noindent
We try our models with different embedding dimensionalities $d$ of latent vectors.
As illustrated in Figure \ref{fig:performance}, both DIR-MF and DIR-RNN perform relatively stable with varying dimensionalities. When the embedding dimensionality is high, both DIR-MF and DIR-RNN tend to be overfitting. 
In Clothing, DIR-MF gets the best performance with $d=50$, while DIR-RNN achieves the best performance when $d=90$. 
In Electronics, the AUC of DIR-MF is the highest when $d$ is 30, while the AUC of DIR-RNN is the highest when $d$ is 70.
We compare two datasets in the same model, and find that clothing needs higher-dimensional vectors to be represented than electronics. 
It may be because that clothing contains more detailed factors than electronics for recommendation. 

\subsubsection{Impact of Implicit Attribute Vector Number}\label{sec:implicit_num}
\begin{figure}[htbp]
\centering

\subfigure[Clothing]{
\begin{minipage}[b]{0.42\textwidth}
\label{fig:cloth_attr} 
\includegraphics[width=1\textwidth]{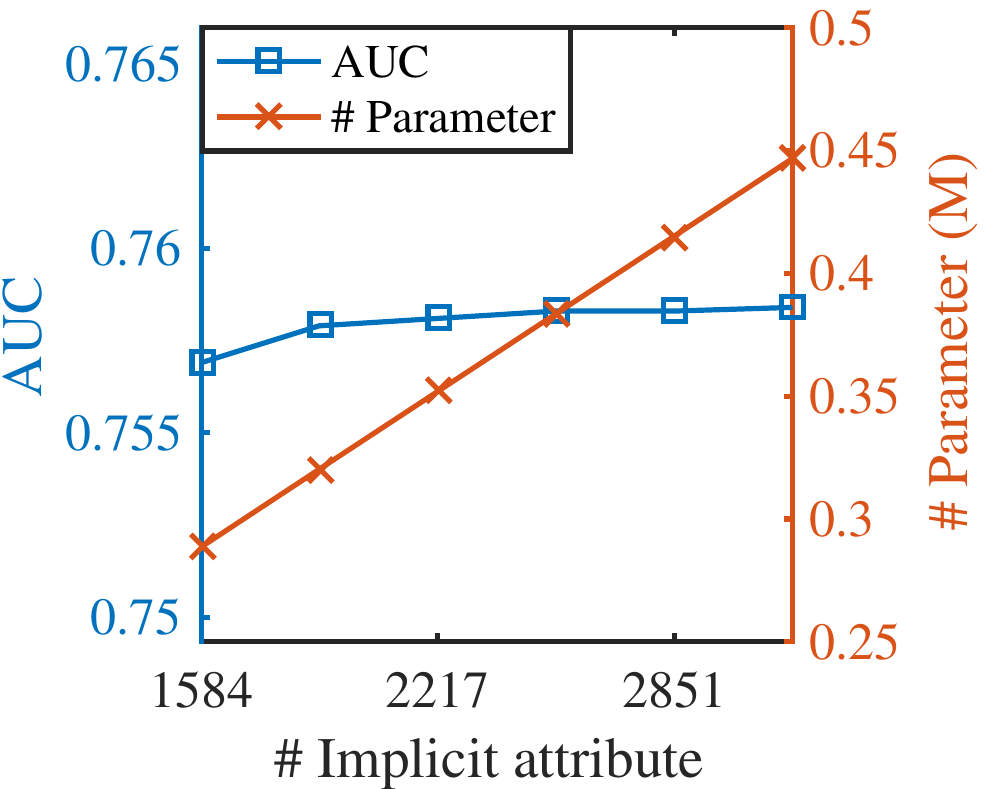}

\end{minipage}%
}%
\subfigure[Electronics]{
\begin{minipage}[b]{0.42\textwidth}
\label{fig:elect_attr} 
\includegraphics[width=1\textwidth]{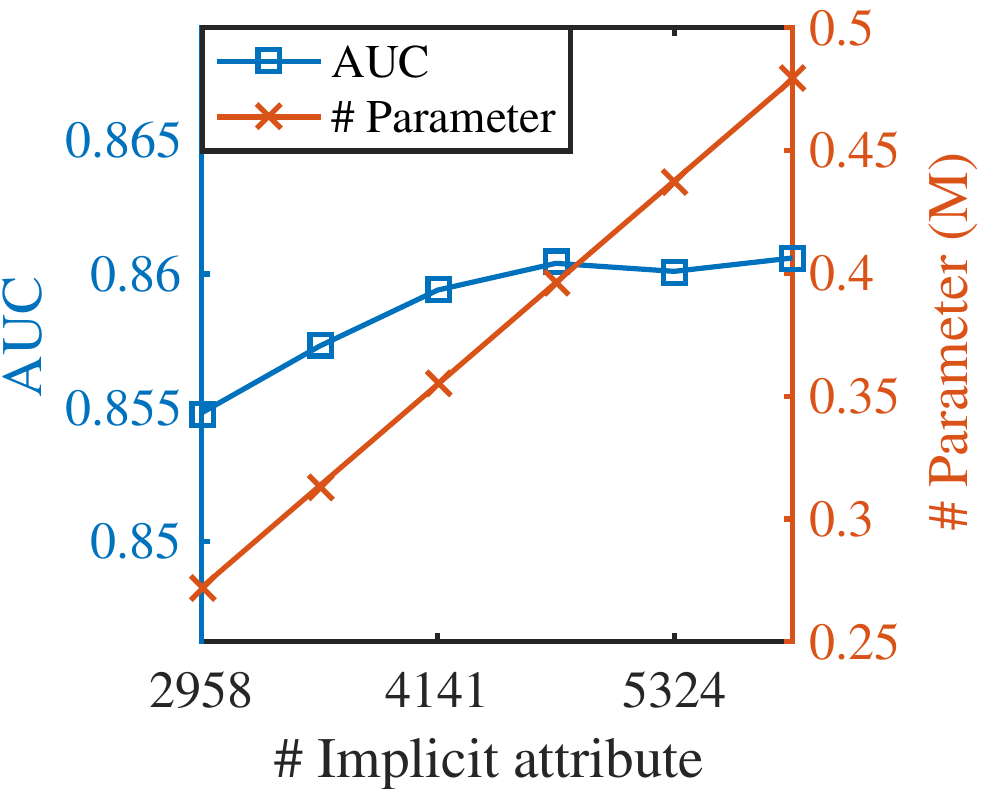}
\end{minipage}%
}%
\caption{AUC performance with different number of implicit attribute on two datasets (Clothing is on the left, Electronics on the right).}
\label{fig:attr_number}
\end{figure}

In the above section, we set the implicit attribute number as 1584 in the Clothing dataset, and  2958 in the Electronics dataset, which is the minimum number could be set. The number of implicit attribute is set to guarantee that every item has different representation combination. 
For example, there are three items in the same category $e^{1}_{1}$. When we represent them by one explicit attribute (category) and one implicit attribute, we need at least three implicit vectors to distinguish them as $(e^{1}_{1}, i_{1})$, $(e^{1}, i_{2})$, $(e^{1}, i_{3})$.
In the Clothing dataset, 1584 is the number of items under the largest explicit attribute (category tag).
In this section, we discuss how DIR performs with the number of implicit attribute number increasing. Figure \ref{fig:attr_number} shows the changes of AUC and parameter numbers along with the increasing of attribute number from the minimum number to $2\times$ minimum number. In the figure, the performance increases at the beginning and becomes stable, when the attribute number increases to $1.8\times$ of the minimum number. The result is reasonable, since when the number of impact increases, the item could get  more implicit vectors to choose. For example, man's clothing and woman clothing may not share some implicit attributes. However, too many implicit attributes is unnecessary.

\subsubsection{Three-attribute item representation}\label{sec:more_attribute}
In the above section, we mainly discuss on only one explicit attribute, i.e., category. In this section, we try to add another attribute (i.e., price) to form a three-attribute representation as category, price, implicit attribute.
In common scene of e-commercial platform, there are plenty of attribute tags except category. However, few datasets could cover all of those information.
In the dataset of Amazon, although there are still some other attribute tags such as price, price. These tags only cover a part of items. We use the price tag as the additional attribute. To have an evenly separation of price, we map the prices which are higher than 5 by the function $5 log_{5} x$. Then, 
we equally divide the range into 5 shares.
So we represent price correspondingly by 5 vectors and an additional vector for the items who do not have price tag in the dataset.

The result is listed in Table \ref{tab:3attr}. By using 3-attribute representation, the parameter space is consequently cut down. However, it does not have great difference compared with 2-attribute representation. The reason is that the different categories tend to have different price range. The minimum number of style vectors (i.e., the largest number of items in the same combination of category and price tag) is $1224$ in the Clothing dataset and $1556$ in the Electronics. The number of style vectors does not change a lot compared with 2-attribute representation i.e., $1584$  in the Clothing dataset and $2958$ in the Electronics. 
In the dataset of Clothing, an additional attribute seems to have side effect on the performance, while in the Electronics dataset the performance gets a little improvement. The main problem of the Clothing dataset comes from too much missing information of item price. There are nearly one-third of items without price tag in the Clothing dataset and only one-tenth of items in the Electronics dataset. This insight gives us an useful conclusion that the information of explicit attributes needs to be accurate. 
The missing or false information of explicit attributes leads the items with different real-world attributes share the same representation part, which confuses DIR to learn the real attributes representation. In that case, under the framework of DIR, it is not good to add an explicit attribute, whose tag is unknown to too much items.

\begin{table*}[htbp]
\small
\centering
\caption{The performance of 3-attribute (price, category, implicit) DIR compared with 2-attribute DIR (category, implicit). *(3) means corresponding model using 3 attributes and *(2) means corresponding model using 2 attributes.}
\vspace{1mm}
    \begin{tabular}{c|c|cc|cc}
    \toprule
    Dataset & Setting & DIR-MF (3) & DIR-MF (2) & DIR-RNN (3)  & DIR-RNN (2)\\

    \midrule
\multirow{3}[0]{*}{Electronics} & warm-start &
0.7644 & 0.7585 & 0.8568  & 0.8548  \\
& \# parameter (M)& 
0.7036 & 0.7594  & 0.1743  & 0.2720 \\
&  time (s) & 
 0.2489 & 0.2379 & 65.3184  & 59.1175\\

    \midrule
\multirow{3}[0]{*}{Clothing} & warm-start & 
   0.6139 & 0.6549 & 0.7094  & 0.7569 \\

& \# parameter (M) & 
 1.1197 & 1.1373 & 0.2534  & 0.2886 \\
& time (s) &
 0.2845 & 0.2738 & 190.0021  & 178.1951 \\

    \bottomrule
    \end{tabular}%
\label{tab:3attr}%
\end{table*}%

\subsubsection{Convergence}\label{sec:convergence}
\begin{figure}[t]
\centering

\includegraphics[width=0.65\textwidth]{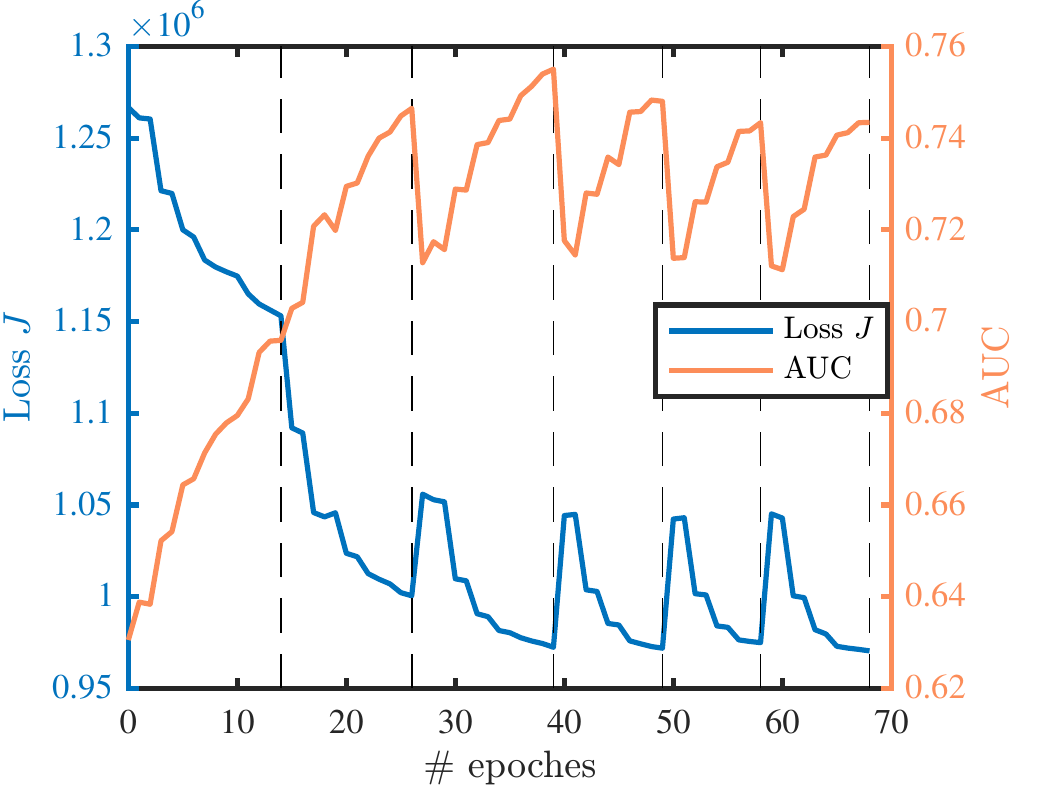}
\caption{The training loss and AUC of DIR during the training epoch. The x-coordinate is the number of training epoch. The dotted bar is the flag of reallocation.}
\label{fig:convergence}

\end{figure}

 \noindent
We look at the loss $J$ during our learning process. In Figure \ref{fig:convergence}, the horizontal ordinate means the number of training epochs. In an epoch, the model adapts all user-item pairs in training set once. The blue curve represents the loss $J$ and the orange curve represents the AUC in valid dataset. The dotted bar is the time when a reallocation is conducted. DIR nearly reaches convergence after 3 reallocations. Although the training process does not totally converge. DIR could allocate the items to a better representations and get a better estimation.
There are reasonable waves after 2 reallocations, since reallocation changes relation of representations which is trained by SGD. 
According to experimental experience, our model tends to achieve the best result in the testing set, after three allocation. 

\subsubsection{Failed examples}
\begin{figure}[t]
\centering

\includegraphics[width=0.7\textwidth]{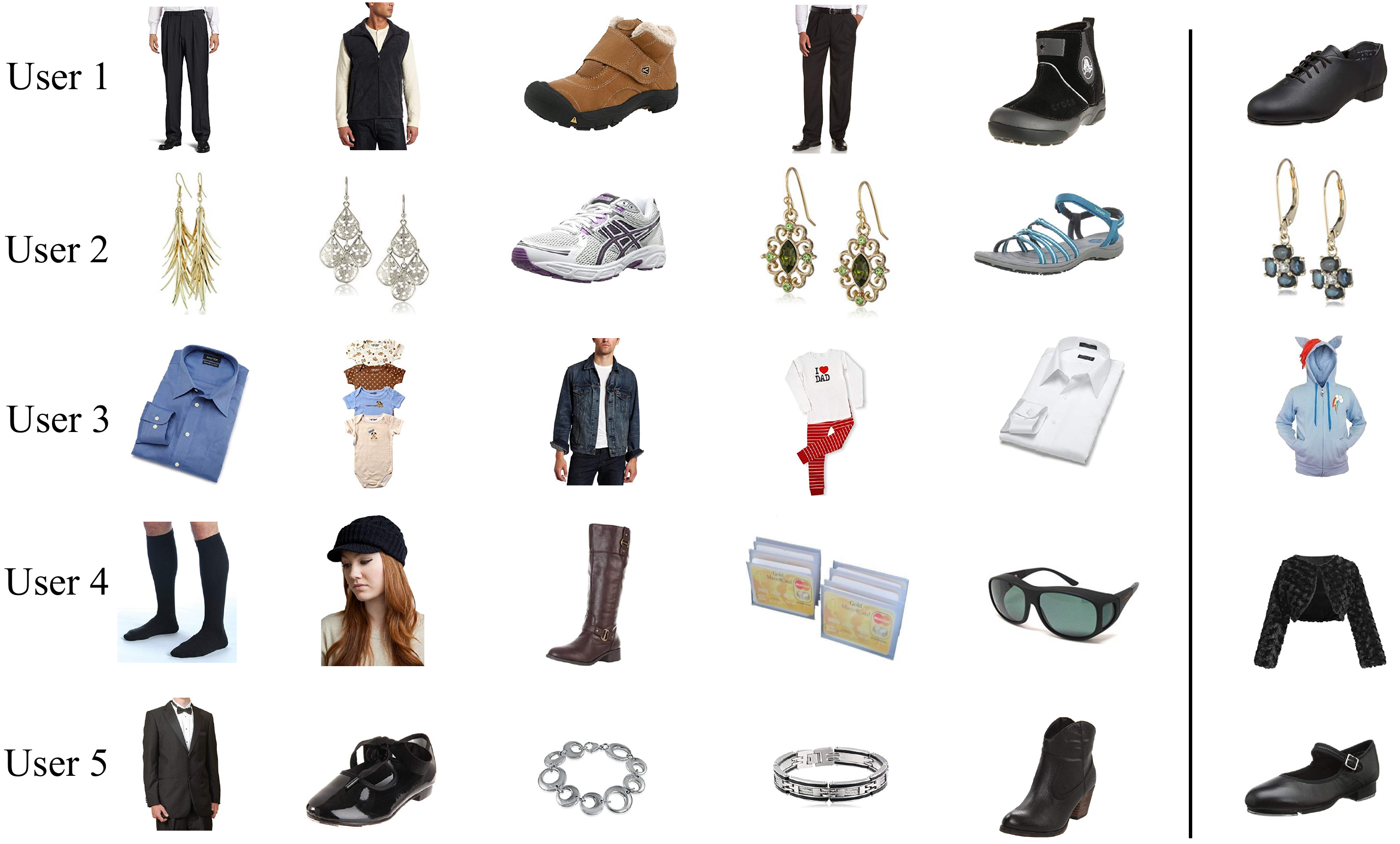}
\caption{Five failed examples of DIR-RNN. Each row is a user's purchase sequence. DIR-RNN is used to predict the last item. The prediction is listed on the right side.}
\label{fig:failed_example}

\end{figure}

In this section, we  list some failed examples of DIR-RNN and analyze the reason in detail. Figure \ref{fig:failed_example}  performs Five failed examples of DIR-RNN. Each row is a user's purchase sequence. The predictions of DIR-RNN are listed on the right side. 
The failed examples mainly come from three reasons. First, DIR-RNN some times could not totally figure out preference changing from one attribute to another. For example, 
User1's purchase history indicts that he prefers formal suits, so DIR-RNN recommends a leather shoes, while the user prefers casual shoes now. User2 always buys ear rings, but her preference changes to shoes recently. 
Second, some users' purchase sequences are difficult to predict, such as User3 and User4. User3 may be a father. He buy things for his child and himself together. DIR-RNN confuses of his preference on categories. User4 tends to buy things of similar style. DIR-RNN recommends a jacket corresponding to the style, but she buy a pair of sunglasses which is also suitable. 
Third, DIR-RNN sometimes ignores the meticulous diversity of items. DIR-RNN recommends a pair of women's leather shoes for User5, which appeal to her preference. As we can see in Figure \ref{fig:failed_example}, User5 has bought a similar pair of leather shoes before, so she chooses a pair of different ones. This difference could only be distinguished by image, which is not yet adopted in DIR-RNN.

\section{Conclusion}
 \noindent
In this paper, we have proposed a novel DIR to model attribute-level disentangled representations of items. The LearnDIR is further introduced to train the models under the framework of DIR.  We apply DIR to two typical models as DIR-MF and DIR-RNN. Our experiments have shown that the DIR could establish an elaborate item representation for collaborative models, and the proposed model outperforms the state-of-the-art results.
DIR allocates the items into tensor-like attribute embeddings, which has great benefits in reducing the parameter size and alleviating the cold-start problem. Besides, the only trade off comes from the complexity during training. Since additional training time does not have much influence in most  real-world applications, DIR is workable.

In the future, first, we are going to take multimodal information like visual information into account, to enrich the representation. Second, we try to adapt the DIR to user representation. 
The code will soon be released \footnote{https://github.com/CRIPAC-DIG/DIR.}.

\begin{acks}
This work is supported by National Key Research and Development Program (2018YFB1402605, 2018YFB1402600), National Natural Science Foundation of China (U19B2038, 61772528), Beijing National Natural Science Foundation (4182066).



\end{acks}

\bibliographystyle{ACM-Reference-Format}
\bibliography{sample-bibliography}

\end{document}